\documentclass[10pt, conference]{IEEEtran}
\IEEEoverridecommandlockouts
\usepackage{cite}
\usepackage{amsmath,amsfonts}
\usepackage{xcolor}
\usepackage[inline,shortlabels]{enumitem}
\usepackage{xspace}
\usepackage{multirow}

\usepackage[ruled,linesnumbered]{algorithm2e}
\usepackage[caption=false,font=footnotesize]{subfig}
\usepackage{inputenc}
\usepackage{mathrsfs}
\usepackage{multirow}
\usepackage{pgfplots}
\usepackage{xspace}
\usepackage{tcolorbox}
\usepackage{graphicx}
\usepackage{lettrine}
\usepackage{pifont}
\usepackage{fancyhdr}
\usepackage{tabularx}
\usepackage{makecell}
\usepackage{booktabs}
\usepackage{calc} 

\usepackage{diagbox}

\usepackage{hyperref}

\usepackage{flushend}

\graphicspath{{figures/}}

\newlength{\lengthc}
\definecolor{ccr}{RGB}{120,29,125}  
\hypersetup{hypertex=true,
    colorlinks=false,
    linkcolor=ccr,
    anchorcolor=ccr,
    pdfborder={0 0 0},    
    citecolor=ccr}

\def\ie{\textit{i.e.},~}

\def\eg{\textit{e.g.},~}

\def\namerm{TrioXpert}
\def\name{\textit{\namerm}\xspace}

\newlist{questions}{enumerate}{1}
\setlist[questions]{wide=0pt, label=\textbf{RQ\arabic*.}, ref=RQ\arabic*}

\def\BibTeX{{\rm B\kern-.05em{\sc i\kern-.025em b}\kern-.08em
    T\kern-.1667em\lower.7ex\hbox{E}\kern-.125emX}}

\begin{document}

\title{TrioXpert: An Automated Incident Management Framework for Microservice System
}

\author{
\IEEEauthorblockN{
Yongqian~Sun\IEEEauthorrefmark{2}\IEEEauthorrefmark{9},
Yu~Luo\IEEEauthorrefmark{2},
Xidao~Wen\IEEEauthorrefmark{5}\thanks{\IEEEauthorrefmark{5}Xidao Wen is the corresponding author.},
Yuan~Yuan\IEEEauthorrefmark{7},
Xiaohui~Nie\IEEEauthorrefmark{3},
Shenglin~Zhang\IEEEauthorrefmark{2}\IEEEauthorrefmark{6} 
Tong~Liu\IEEEauthorrefmark{4},
Xi~Luo\IEEEauthorrefmark{4}
}

\IEEEauthorblockA{
\IEEEauthorrefmark{2}Nankai~University, 
\{sunyongqian, zhangsl\}@nankai.edu.cn,
\{luoyu\}@mail.nankai.edu.cn \\
\IEEEauthorrefmark{3}Computer~Network~Information~Center,~Chinese~Academy~of~Sciences, xhnie@cnic.cn	 \\
\IEEEauthorrefmark{4}Lenovo (TianJin) Co., Ltd., 
\{liutong14, luoxi9\}@lenovo.com \\
\IEEEauthorrefmark{5}BizSeer, wenxidao@bizseer.com \\
\IEEEauthorrefmark{6}Key Laboratory of Data and Intelligent System Security, Ministry of Education, China \\
\IEEEauthorrefmark{7}National University of Defense Technology, yuanyuan@nudt.edu.cn \\
\IEEEauthorrefmark{9}Tianjin Key Laboratory of Software Experience and Human Computer Interaction \\
}

}

\maketitle
\thispagestyle{fancy}
\renewcommand{\headrulewidth}{0pt}
\renewcommand{\footrulewidth}{0pt}
\pagestyle{fancy}
\cfoot{\thepage}

\begin{abstract}
Automated incident management plays a pivotal role in large-scale microservice systems. 
However, many existing methods rely solely on single-modal data (\eg metrics, logs, and traces) and struggle to simultaneously address multiple downstream tasks, including anomaly detection (AD), failure triage (FT), and root cause localization (RCL).
Moreover, the lack of clear reasoning evidence in current techniques often leads to insufficient interpretability.
To address these limitations, we propose \name, an end-to-end incident management framework capable of fully leveraging multimodal data. 
\name designs three independent data processing pipelines based on the inherent characteristics of different modalities, comprehensively characterizing the operational status of microservice systems from both numerical and textual dimensions.
It employs a collaborative reasoning mechanism using large language models (LLMs) to simultaneously handle multiple tasks while providing clear reasoning evidence to ensure strong interpretability.
We conducted extensive evaluations on two microservice system datasets, and the experimental results demonstrate that \name achieves outstanding performance in AD (improving by 4.7\% to 57.7\%), FT (improving by 2.1\% to 40.6\%), and RCL (improving by 1.6\% to 163.1\%) tasks.
\name has also been deployed in Lenovo's production environment, demonstrating substantial gains in diagnostic efficiency and accuracy.
\end{abstract}

\begin{IEEEkeywords}
Anomaly Detection, Failure Triage, Root Cause Localization, Microservice System
\end{IEEEkeywords}

\section{Introduction}\label{intro}

Microservice architectures have become the standard for modern enterprise systems, offering scalability and modular deployment \cite{art, diagfusion}. However, their dynamic behavior and complex interdependencies \cite{automap, eadro} pose significant challenges for operations. Localized faults can quickly cascade, resulting in system-wide outages and business disruption.
Timely and accurate incident management is thus essential. Traditional approaches rely on on-call engineers (OCEs) to manually inspect metrics, logs, and traces—a process increasingly impractical at scale.
In response, AIOps has emerged to automate incident response using data-driven and learning-based techniques.

\begin{figure}[t]
    \centering
    \includegraphics[width=\linewidth]{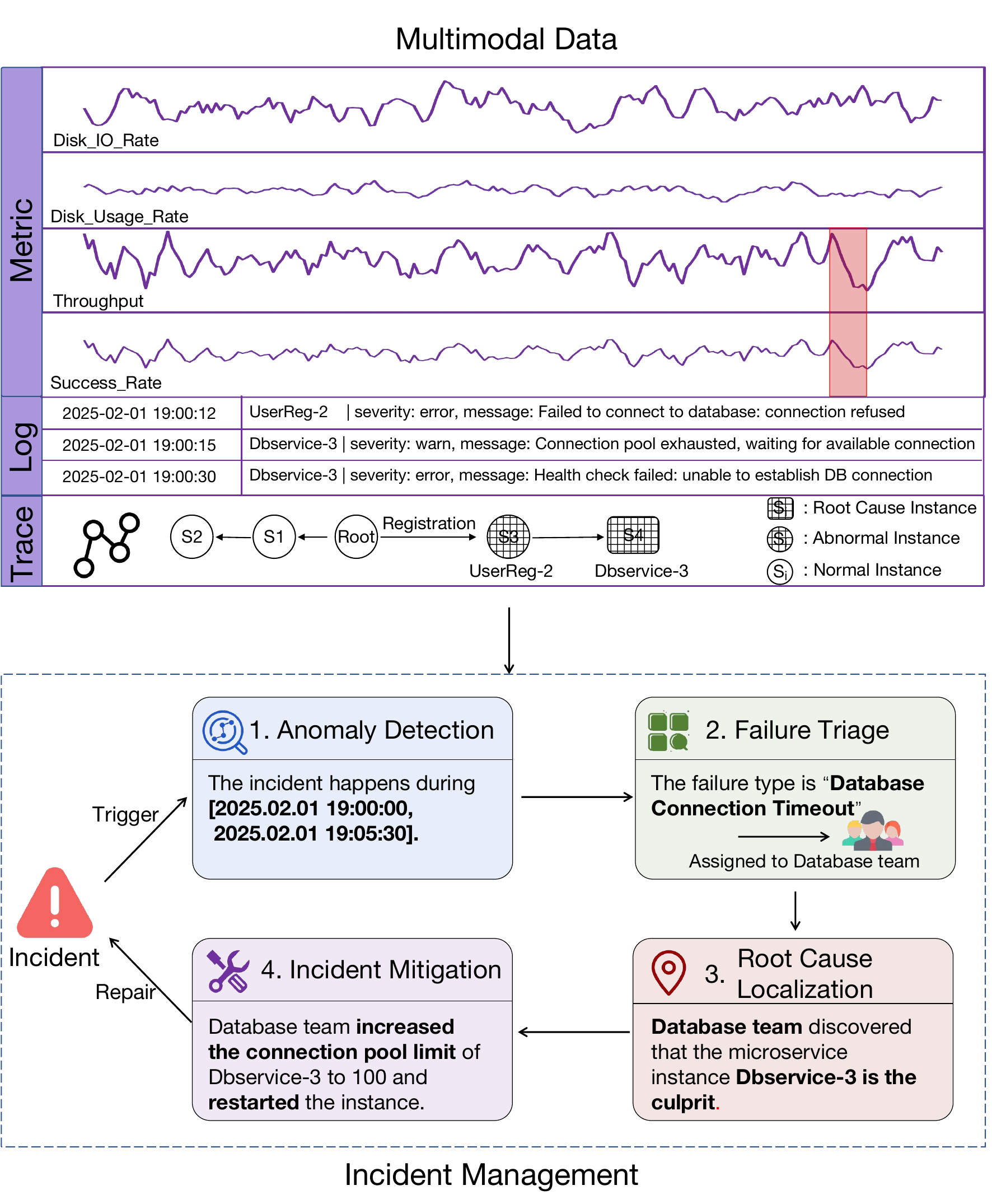}
    \caption{
        Multimodal data and incident lifecycle. (Top) Examples of the three data modalities used in incident management – metrics (time-series signals), logs (timestamped events), and a trace (service call graph). (Bottom) The four-stage incident management lifecycle: (1) \underline{A}nomaly \underline{D}etection, (2) \underline{F}ailure \underline{T}riage, (3) \underline{R}oot \underline{C}ause \underline{L}ocalization, and (4) \underline{I}ncident \underline{M}itigation.
        }
    \label{fig:workflow_multimodal}
    \vspace{-8mm}
\end{figure}

As shown in Fig.~\ref{fig:workflow_multimodal}, the typical lifecycle of software incident management is divided into four stages \cite{rcacopilot, art, detection_better}: 
(1) \textbf{Anomaly Detection (AD)}: Initially, an alert is generated upon detecting deviations in system behavior, triggering the incident management process.
(2) \textbf{Failure Triage (FT)}: The incident is then categorized based on its characteristics and assigned to the appropriate team.  
(3) \textbf{Root Cause Localization (RCL)}: A comprehensive analysis of all aspects of the incident is conducted to identify the root cause.
(4) \textbf{Incident Mitigation (IM)}: OCEs take appropriate measures based on the preceding diagnostic results to restore system functionality.

Among the four stages, AD, FT, and RCL are both critical and amenable to automation, making them the primary targets for AI-driven solutions. In contrast, mitigation typically involves manual intervention and complex orchestration, thus remains out of scope for this work. We therefore focus on advancing automation across the first three stages.

Over the past decade, numerous AIOps models have been proposed to enable automated incident management. 
Although these approaches have demonstrated strong performance in specific scenarios, their practical application has gradually revealed several critical limitations as software systems continue to grow in scale and complexity.

A major limitation lies in the fragmented design of existing methods: most models are developed for single subtasks \cite{loganomaly,traceanomaly,isquad,arvalus,microcbr,unidiag,microrank,coral,tracerca,microsketch}, rather than end-to-end workflows. This leads to increased deployment costs and integration overhead in real-world environments~\cite{art}. 
To address this issue, a few recent studies attempt to unify multiple tasks \cite{art, eadro, shapleyiq, dejavu, diagfusion, zhou2019latent}. However, many of these approaches still face notable shortcomings: some approaches utilize supervised learning paradigms that demand high-quality labels~\cite{eadro,dejavu, diagfusion, zhou2019latent}, while others require domain experts to handcraft rules or causal graphs~\cite{shapleyiq}. Among them, ART~\cite{art} stands out by jointly modeling multiple tasks through self-supervised learning, requiring only unlabeled data and minimal human effort. Moreover, it attempts to integrate heterogeneous multimodal data into a unified representation, offering a promising foundation for holistic incident management.

However, the multimodal fusion in ART still faces two key challenges. First, it reduces logs and traces to superficial statistical features (\ie log template frequencies, trace durations), neglecting rich textual semantics that are essential for accurate failure understanding. Second, the sheer volume of logs and traces often exceeds processing capacity, introducing noise and computational bottlenecks that hinder efficient analysis.

Beyond modeling limitations, incident management also demands high interpretability. 
Interpretability, in our context, refers to the ability to provide a reasonable and logically rigorous reasoning process for diagnostic results, supported by data-driven analysis. 
While deep-learning-based methods excel in accuracy, their ``black-box'' nature makes it difficult for OCEs to understand and trust model outputs. 
Although systems like ART~\cite{art} enhance interpretability through deviation-based heuristics, they still lack structured, step-by-step reasoning that OCEs can inspect or validate.

In recent years, the rapid advancement of large language models (LLMs) has demonstrated remarkable capabilities in complex reasoning tasks such as math and programming, making them promising candidates for incident management.  
Moreover, their ability to generate natural language explanations aligns well with the interpretability requirements in incident management.
However, current research applying LLMs to incident management has not adequately addressed the aforementioned challenges.
For instance, works such as Oasis\cite{oasis}, XPERT \cite{xpert}, and NISSIST\cite{nissist} primarily focus on isolated subtasks (\ie summary generation, KQL query construction, and mitigation recommendation), while COMET \cite{comet} and LasRCA\cite{lasrca} are limited to single-task failure triage or root cause localization.
Moreover, many recent studies \cite{rcacopilot, icl_rca, llm4rca, loglm} rely solely on logs or partial diagnostic data, lacking comprehensive integration of metrics, logs, and traces.
As a result, existing frameworks fail to support unified, multimodal, and multi-task analysis across the full incident management lifecycle—including anomaly detection, failure triage, and root cause localization.

Even if we design a better framework integrating LLMs to address these limitations, intrinsic constraints of LLMs pose the third challenge to its effectiveness in real-world deployment.
Internally, issues such as hallucinations and context window limits are problems that persist regardless of model prompting or fine-tuning strategies.  
These limitations further prevent models from generating stable, rigorous reasoning paths that satisfy the interpretability requirements of incident management.
We summarize the challenges as follows, and more details are provided in Section~\ref{background}:
\begin{enumerate}[leftmargin=1.5em]
\item \makebox[\lengthc][r]{\textbf{Challenge~1:}} Semantic impoverishment in multimodal fusion.
\item \makebox[\lengthc][r]{\textbf{Challenge~2:}} Textual data overload in real-time incident management.
\item \makebox[\lengthc][r]{\textbf{Challenge~3:}} LLM limitations in complex and trust-critical incident management.
\end{enumerate}

To tackle the aforementioned challenges, this paper proposes \name, an end-to-end incident management framework based on LLMs collaboration.
Specifically, \name consists of three core modules:
(1) \textit{Multimodal Data Preprocessing}: To handle the heterogeneity of metrics, logs, and traces, \name employs three modality-specific preprocessing pipelines and two filtering mechanisms to extract incident-relevant logs and traces, reducing noise and improving data quality.
(2) \textit{Multi-Dimensional System Status Representation}: This module consolidates numerical and textual features into a unified multi-view representation, capturing statistical patterns, semantic context, and service interactions.
(3) \textit{LLMs Collaborative Reasoning}: An multi-expert collaboration architecture coordinates three LLM-based agents—Numerical, Textual, and Incident Experts—to jointly perform AD, FT, and RCL. Outputs include structured reasoning chains to enhance interpretability and trust.
The contributions of this work are summarized as follows:

\begin{enumerate}[leftmargin=1.5em]

\item We propose \name, the first end-to-end framework that unifies AD, FT, and RCL using a collaborative reasoning architecture built on LLMs. The design integrates metrics, logs, and traces through a structured prompting strategy, enabling scalable and explainable multi-task diagnostics.
\item \name incorporates dedicated pipelines and filtering mechanisms to extract and align heterogeneous observability data. This enables a more complete system state representation from both numerical and textual perspectives, improving both diagnostic accuracy and interpretability.
\item We evaluated \name on two real-world microservice systems and observed consistent improvements over baselines in both performance and interpretability. \name has also been deployed in Lenovo's production environment, demonstrating substantial gains in diagnostic efficiency and accuracy. To ensure reproducibility, we release all code, prompts, configurations, and data~\footnote{https://anonymous.4open.science/r/TrioXpert-F244}.
\end{enumerate}

\section{Background}\label{background}

\subsection{Motivating Study}
\subsubsection{Do Metrics, Logs, and Traces All Carry Diagnostic Value}


To understand how various modalities characterize microservice systems, we conducted a qualitative and quantitative analysis of operational data collected from Lenovo's production environment. 
Specifically, we collected 20 incident cases and gathered corresponding metrics, raw logs, and traces within a 10-minute window around each incident—reflecting standard incident diagnosis practices in industry deployments \cite{IBM-10}.

Our findings are threefold:
(1) Metrics reflect system health in a quantitative and timely manner, but lack contextual information to identify root causes.
(2) Logs often contain explicit failure clues, such as error or warning messages like ``Connection refused'' or ``Timeout exceeded'', which are highly informative for diagnosis. However, the vast majority of log entries are routine debug or info-level messages that are unrelated to incidents (\eg \textit{``severity: info, message: conversion request successful''}).
(3) Similarly, traces capture detailed execution paths across microservices and can reveal dependency anomalies. However, most traces correspond to normal operations, diluting their usefulness in root cause analysis. For instance, during a database timeout incident, only 12 out of more than 8,000 trace spans reflected abnormal behavior.

\begin{center}
\begin{tcolorbox}[colback=gray!10,
                  colframe=black,
                  width=8.5cm,
                  arc=5mm, auto outer arc,
                  boxrule=0.5pt,
                 ]
\textbf{Takeaway 1: Metrics, logs, and traces reflect distinct aspects of system behavior, all of which contribute valuable diagnostic signals. However, logs and traces are often dominated by redundant entries that hinder efficient analysis.}
\end{tcolorbox}
\end{center}

\subsubsection{Can a Single LLM Reliably Perform Well over Multimodal Inputs in Incident Management}

To assess the reliability of a single LLM in multimodal incident analysis, we conducted an exploratory study on the same 20 cases used in the previous experiments. The model received concatenated inputs comprising metrics, logs, and traces, and was prompted to generate diagnostic reports addressing three tasks: AD, FT, and RCL. The outputs were subsequently reviewed by two experienced OCEs.

While the LLM often produced fluent and seemingly structured responses, both OCEs noted that:
(1) In 50\% of the cases, the diagnostic conclusions were factually incorrect.
(2) Even when plausible, the reasoning paths were frequently untraceable, with fabricated intermediate steps that impeded logical validation.
(3) The model struggled with incidents involving long service dependency chains or subtle textual anomalies, as its fixed context window constrained the amount of usable input and often resulted in the loss of diagnostic signals through truncation.

These findings underscore the limitations of using a monolithic LLM for end-to-end incident management. Hallucinations, opaque reasoning, and context window constraints significantly impair both the accuracy and trustworthiness of the outputs in realistic settings.

\begin{center}
\begin{tcolorbox}[colback=gray!10,
                  colframe=black,
                  width=8.5cm,
                  arc=5mm, auto outer arc,
                  boxrule=0.5pt,
                 ]
\textbf{Takeaway 2: A single LLM often fails to produce reliable and interpretable results when directly applied to complex incident management tasks.}
\end{tcolorbox}
\end{center}

\subsection{Challenges}

Based on the above motivating studies, we summarize the following key challenges:
\subsubsection{Semantic Impoverishment in Multimodal Fusion}
Although recent multimodal AIOps approaches~\cite{art, diagfusion, eadro} have achieved promising results on downstream tasks such as AD, FT, and RCL, they often fall short in capturing the full diagnostic value of each modality.  

For metrics, modeling based on time-series trends is both appropriate and effective. 
However, for logs, these methods typically extract only superficial statistical features—such as log template frequencies~\cite{art} or event-template IDs~\cite{diagfusion}—ignoring rich textual semantics that are critical for failure diagnosis. 
A representative example is an authentication service failure caused by expired credentials.  
In this case, existing methods detected a high frequency of a specific event type and mild latency deviations but failed to identify the root cause. 
In contrast, a critical log entry (\ie \textit{``message: authentication failed due to expired token''}) directly revealed the root cause, which is an insight lost under purely numerical modeling.  

Similarly, trace data is often reduced to coarse-grained statistics like service dependency graphs and request durations.
This approach overlooks fine-grained signals embedded in traces, such as call types and status codes, which can reveal subtle yet important failure patterns.  

As a result, current fusion strategies fail to align heterogeneous modalities at a meaningful semantic level, leading to incomplete system state characterization and unreliable diagnostic outcomes.


\subsubsection{Textual Data Overload in Real-Time Incident Management}

While logs and traces contain rich diagnostic semantics, their sheer volume poses a major obstacle to effective processing.  
In even the most minor incidents, logs and traces often generate over 10,000 entries independently.  
This scale not only incurs high computational overhead but also obscures the identification of incident-relevant signals.  

Neither manual inspection nor basic natural language processing techniques can efficiently process such massive volumes in time-sensitive incident management scenarios.  
Furthermore, the vast majority of log and trace entries are routine or irrelevant to incidents, offering little diagnostic value while introducing significant noise into the analysis.  
These issues highlight the need for effective filtering mechanisms that distill actionable signals from logs and traces before reasoning.

\subsubsection{LLM Limitations in Complex and Trust-Critical Incident Management}


While LLMs offer strong general reasoning capabilities, their application to incident management is fundamentally constrained by architectural limitations that hinder the generation of reliable and interpretable diagnostic outputs.
Specifically, long-context reasoning is impeded by token limits, which truncate critical signals from extended telemetry sequences—particularly problematic when analyzing cascading failures across microservices.  
More critically, hallucinations (\eg fabricated service dependencies or spurious causal explanations) introduce significant risks in high-stakes environments, where diagnostic errors can delay mitigation or erode operational trust.
These issues hinder the system's ability to produce explicit, logically coherent reasoning paths that align with real-world failure patterns.  

\subsection{Problem Definition}\label{problem}
Typically, OCEs need to collect multimodal data (\ie metrics, logs, and traces) and perform three core tasks: AD, FT, and RCL.
Specifically, the goal of the AD task is to detect whether there are multiple abnormal timestamps within a time window, thereby determining whether the system has encountered an anomaly. 
The FT task involves selecting the most probable failure type $c_i$ from a predefined set of fault categories $\mathbb{C}=\{c_1,c_2,...,c_n\}$.
Finally, the RCL task identifies the most likely root cause $r_i$ from all service instances $\mathbb{R}=\{r_1,r_2,...,r_m\}$.

However, traditional frameworks execute these tasks in a fixed sequence, lacking the flexibility to adapt to real-world operational needs. 
To address this, \name leverages LLMs to support customized task execution. 
Given a time window $T$ and a subset of tasks $\mathbb{S}={AD, FT, RCL}$, \name constructs a unified system representation and completes the selected tasks via LLM-based reasoning.

\section{Methodology}
\subsection{Overview}

\begin{figure*}[t]
    \centering
    \includegraphics[width=\linewidth]{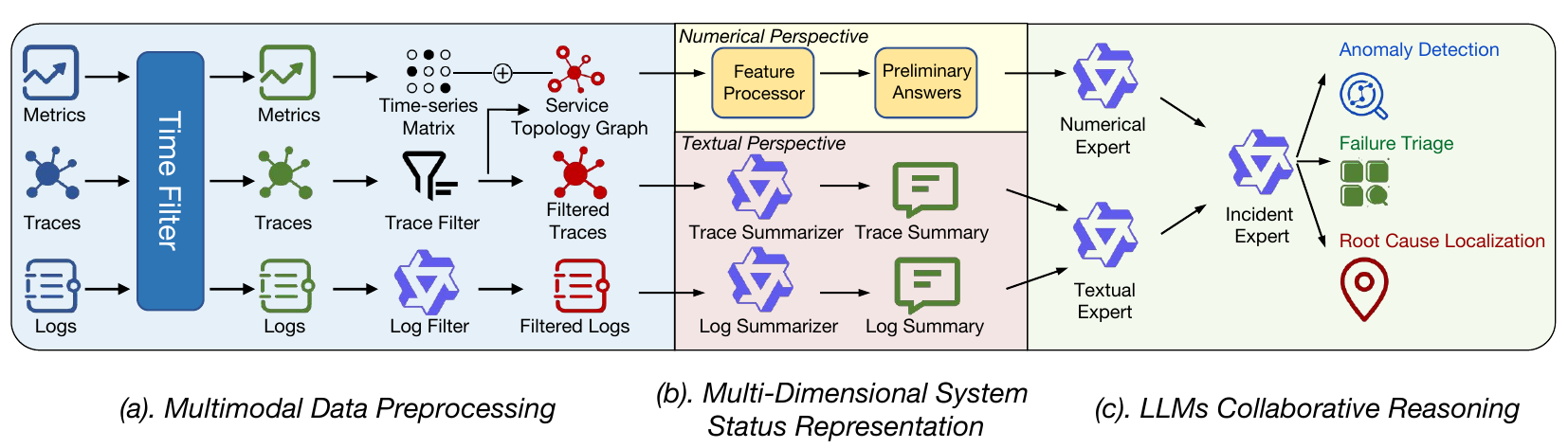}
    \caption{
        The overview of \name.  The framework consists of three modules: (a) Multimodal Data Preprocessing, which filters and prepares metrics, traces, and logs (e.g., aligning by time and extracting relevant subsets); (b) Multi-Dimensional System Status Representation, which derives a numerical feature-based preliminary analysis and textual summaries of logs/traces (via specialized “Log Summarizer” and “Trace Summarizer”); and (c) LLMs Collaborative Reasoning, an multi-expert collaboration architecture comprising a Numerical Expert and Textual Expert that collaborate with an Incident Expert LLM to perform anomaly detection, failure triage, and root cause localization simultaneously.
    }
    \label{fig:overview}
    \vspace{-4mm}
\end{figure*}

As illustrated in Fig.~\ref{fig:overview}, \name comprises three core modules: (1) \textit{Multimodal Data Preprocessing}, (2) \textit{Multi-Dimensional System Status Representation}, and (3) \textit{LLMs Collaborative Reasoning}. These modules are orchestrated to achieve accurate and interpretable incident analysis.

In the preprocessing stage, tailored strategies are applied to metrics, logs, and traces to produce structured inputs: a time-series matrix $\mathcal{M}$, a service topology graph $\mathcal{G}$, and filtered logs $\mathcal{L}$ and traces $\mathcal{T}$. Subsequently, three modality-aware pipelines extract numerical and textual features for system status characterization. Finally, an multi-expert collaboration architecture coordinates multiple LLMs to jointly perform downstream incident management tasks, supported by structured reasoning evidence.

\vspace{-1.8mm}

\subsection{Multimodal Data Preprocessing}

To effectively leverage the rich information contained in multimodal data while minimizing information loss, this module performs specific preprocessing on three modalities of data to adapt them to subsequent three data processing pipelines.

\subsubsection{Metrics}
Considering the time-series characteristics of metrics, it is organized into a three-dimensional time-series matrix $\mathcal{M}$. 
The three dimensions of this matrix represent timestamps, service instances (\eg emailservice-0, logservice-1), and feature channels (\eg cpu\_util, disk\_io\_rate), respectively.
This structured representation not only facilitates subsequent time-series feature extraction but also effectively captures the dynamic changes in system operational status.

\begin{center}
\begin{tcolorbox}[colback=gray!10,
                  colframe=black,
                  width=8.5cm,
                  arc=5mm, auto outer arc,
                  boxrule=0.5pt,
                 ]
\textbf{Definition 1: Time-series matrix} is a three-dimensional tensor $\mathcal{M} \in \mathbb{R}^{T \times S \times F}$ where $T$ represents the number of timestamps in the time series, $S$ denotes the number of service instances, and $F$ indicates the number of feature channels.
\end{tcolorbox}
\end{center}

\subsubsection{Logs}

\name employs an LLM-based two-stage filtering mechanism to isolate incident-relevant log entries, reducing noise and improving cross-scenario adaptability. Inspired by COMET~\cite{comet}, the method preserves the core idea of combining keyword filtering with semantic refinement, but replaces manual heuristics with instruction-driven automation.

In the first stage, the LLM is prompted to extract incident-relevant keywords from the dataset (\eg  generating terms such as \textit{``error'', ``failure'', or ``timeout''} in response to the instruction \textit{``Identify key terms commonly associated with system failures in the following logs''}). 
This approach eliminates the need for labor-intensive manual curation while ensuring comprehensive coverage of incident-related terminology across different system contexts.
The keywords are then employed to filter candidate logs, which are subsequently fed into the second stage for deeper semantic analysis.
In the second stage, the LLM executes context-aware filtering through a structured evaluation prompt.
This instruction-driven mechanism replaces TF-IDF's surface-level term frequency analysis with semantic reasoning, enabling the model to:
(1) identify causally relevant patterns beyond lexical matching,
(2) dynamically weigh diagnostic significance based on contextual evidence,
(3) generalize from few examples without handcrafted thresholds~\cite{loglm}.

Through this process, the filtered logs $\mathcal{L}$ can more accurately indicate incident-related information while eliminating interference caused by irrelevant logs.

\begin{center}
\begin{tcolorbox}[colback=gray!10,
                  colframe=black,
                  width=8.5cm,
                  arc=5mm, auto outer arc,
                  boxrule=0.5pt,
                 ]
\textbf{Definition 2: Filtered logs} $\mathcal{L}$ are a refined subset of log entries obtained through a two-stage process: keyword-based filtering to retain incident-related logs, followed by LLM prioritization to select distinctive entries.
\end{tcolorbox}
\end{center}

\subsubsection{Traces}
To identify anomalous invocation chains while preserving end-to-end causality, we implement a type-aware trace filtering mechanism. 
First, for each invocation type (\eg HTTP, RPC), we compute a latency threshold as the P95 percentile across all spans of that type, accounting for their distinct latency distributions.
Then, when a span's latency exceeds its type-specific threshold, we recursively include all ancestor spans up to the root node. 
This ensures two critical properties: (1) all spans contributing to high-latency paths are retained, and (2) no partial invocation chains are preserved, which is essential for accurate root cause localization.
Additionally, following the methods of ART \cite{art} and DiagFusion \cite{diagfusion}, \name extracts service topology graphs $\mathcal{G}$ from trace data, maintaining the same sampling frequency as the metrics data to describe service instance invocation relationships at any given time.

\begin{center}
\begin{tcolorbox}[colback=gray!10,
                  colframe=black,
                  width=8.5cm,
                  arc=5mm, auto outer arc,
                  boxrule=0.5pt,
                 ]
\textbf{Definition 3: Service topology graph} $\mathcal{G}$ is a directed graph $\mathcal{G}=(V,E)$, where $V$ represents the set of service instances and $E \subseteq V \times V$ denotes the invocation relationships between them.          \\
\textbf{Definition 4: Filtered traces} $\mathcal{T}$ are a subset of trace data containing spans with latency exceeding the P95 threshold, recursively traced back to preserve complete invocation chains indicative of anomalous behavior.
\end{tcolorbox}
\end{center}

Through the above preprocessing, four formalized input datasets are obtained: a 3D time-series matrix $\mathcal{M}$ for metrics, service topology graphs $\mathcal{G}$, filtered logs $\mathcal{L}$, and filtered traces $\mathcal{T}$. 
These inputs support the subsequent data processing pipelines, enabling comprehensive system analysis and efficient reasoning.

\subsection{Multi-Dimensional System Status Representation}
To comprehensively capture the state of a microservice system, \name employs three independent data processing pipelines, targeting both numerical and textual modalities. Each pipeline extracts and encodes salient features from specific modalities, providing structured inputs for downstream reasoning.

\subsubsection{Numerical Perspective}
The numerical pipeline constructs a quantitative system state representation from the time-series matrix $\mathcal{M}$ and service topology graph $\mathcal{G}$.
To model heterogeneous dependencies, we adopt the established feature processor architecture from ART~\cite{art}: a Transformer Encoder captures channel-wise dependencies, a GRU models temporal dynamics, and GraphSage encodes call-level structural relationships.
This processor predicts next-step values of $\mathcal{M}$ and computes deviations from observed metrics.
With the deviations above, we generate task-specific preliminary answers using: (a) extreme value theory thresholding for AD, (b) cut-tree-based clustering for FT, and (c) deviation similarity sorting for RCL. 
All downstream processing follows ART~\cite{art} to ensure methodological consistency with standard diagnostic practices.
Crucially, the reliability of these preliminary answers stems from self-supervised training, which minimizes the L2 norm of prediction errors to learn robust normal-behavior patterns.
Thus, the preliminary answers form a compact numerical representation of the microservice system state.

\subsubsection{Textual Perspective}
\begin{figure}[htbp]
    \centering
    \includegraphics[width=\linewidth]{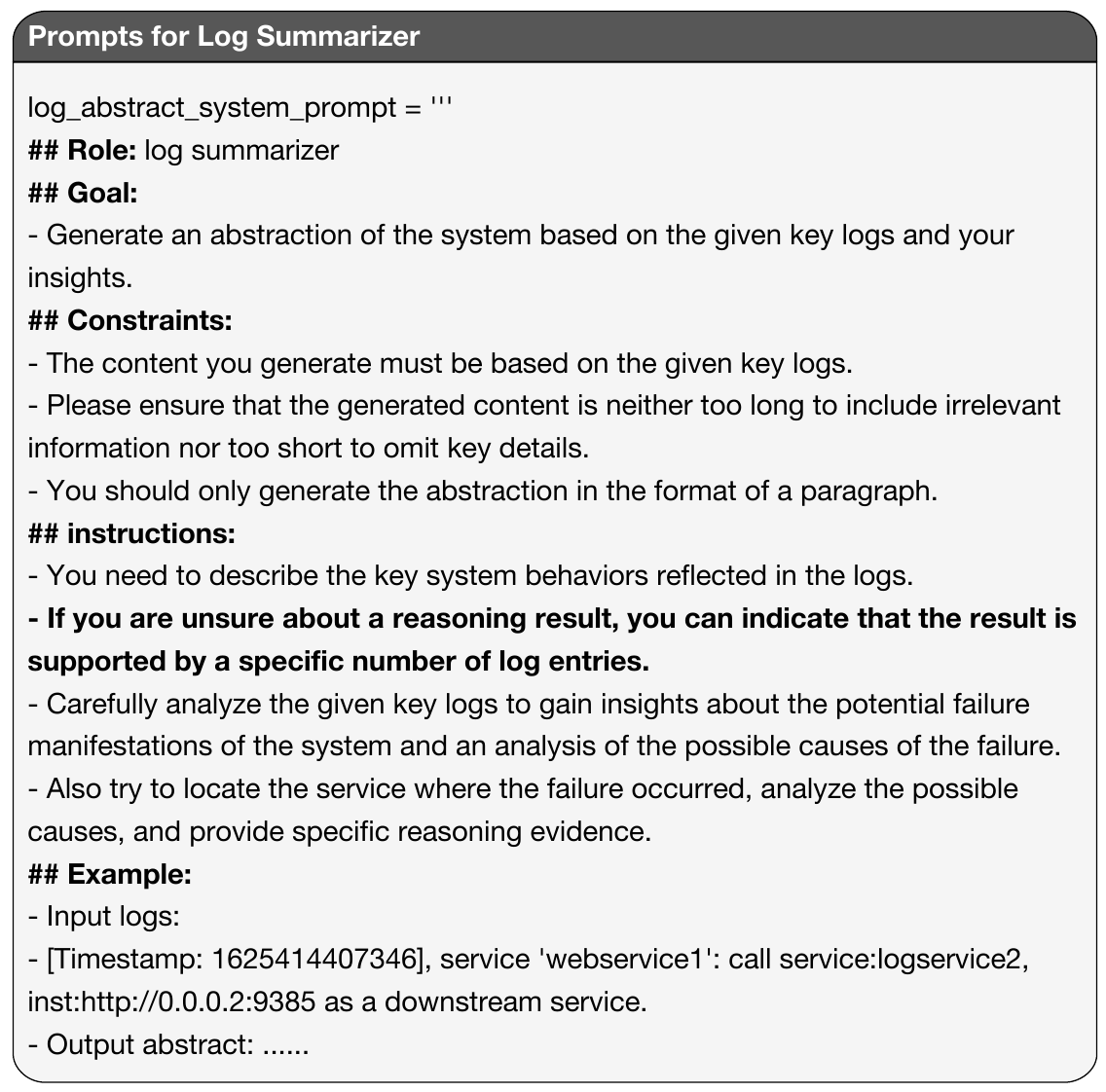}
    \caption{
        The prompt of Log Summarizer.  This prompt utilizes the ``RGCIE'' principle (\ie Role, Goal, Constraints, Instructions, Example) to define the Log Summarizer, which helps mitigate the risk of hallucination.
    }
    \label{fig:prompt_summary}
    \vspace{-5mm}
\end{figure}

The second and third pipelines operate on filtered logs $\mathcal{L}$ and traces $\mathcal{T}$, respectively. Although the preprocessing stage has already removed irrelevant entries, the remaining data can still overwhelm LLMs due to context length constraints. To mitigate this, both pipelines invoke specialized LLMs—Log Summarizer (shown in Fig.~\ref{fig:prompt_summary}) and Trace Summarizer—to summarize their inputs into compact natural language descriptions highlighting abnormal patterns. These summaries serve as low-noise, semantically rich inputs to the reasoning module. 

Through the above three data processing pipelines, \name achieves a comprehensive sketch of the microservice system's status from both numerical and textual dimensions. 
The numerical feature pipeline focuses on capturing the dynamic changes of the system to derive preliminary results, while the textual feature pipelines employ summarization techniques to distill key information, thereby mitigating the interference caused by information overload during the reasoning process.

\subsection{LLMs Collaborative Reasoning}

Despite the notable reasoning capacity of modern LLMs, incident management remains a highly complex, multi-step task requiring precise diagnosis and transparent logic. Relying on a single LLM often results in suboptimal accuracy and poor interpretability due to context overload and hallucinations.

To address these challenges, \name adopts the multi-expert collaboration architecture comprising three specialized LLMs: a Numerical Expert, a Textual Expert, and an Incident Expert. Each expert operates on modality-specific inputs and contributes to a collaborative diagnostic process.
\subsubsection{Definition of LLMs Experts}
\begin{enumerate}[leftmargin=1.5em]
\item \textbf{Numerical Expert}.
The Numerical Expert focuses on analyzing numerical feature representations. 
Its input consists of numerical features $\mathcal{F}_{num}$ generated from time-series matrix $\mathcal{M}$ and service topology graph $\mathcal{G}$, while its output includes the results $\mathcal{R}_{num}$ and corresponding reasoning evidence $\mathcal{E}_{num}$ for three downstream tasks: AD, FT, and RCL.
Specifically, preliminary diagnostic results are generated through a deviation matrix, and the Numerical Expert is responsible for further validating the reasonableness of these results to ensure diagnostic accuracy and reliability, while providing clear explanatory support.
\item \textbf{Textual Expert}.
The Textual Expert is responsible for processing textual feature representations $\mathcal{F}_{txt}$ derived from filtered logs $\mathcal{L}$ and filtered traces $\mathcal{T}$.
Its inputs are log summaries and trace summaries, and its output also includes the results $\mathcal{R}_{txt}$ and corresponding reasoning evidence $\mathcal{R}_{txt}$. 
The objective of the Textual Expert is to analyze textual information to resolve contradictions between data from different sources (\ie logs, traces), extract and emphasize commonalities, and thereby supplement critical details that may be overlooked by numerical features.
\item \textbf{Incident Expert}.
The Incident Expert takes as input the diagnostic results and corresponding reasoning evidence generated by the Numerical Expert and Textual Expert, which originate from different data modalities.
Its output includes the final results $\mathcal{R}_{final}$ and reasoning evidence $\mathcal{E}_{final}$ for AD, FT, and RCL.
Due to potential conflicts or inconsistencies arising from the different origins of numerical and textual features, the Incident Expert employs a comprehensive reasoning mechanism to balance information from various sources, eliminate contradictions, and generate highly consistent and reliable diagnostic results.
\end{enumerate}

\begin{figure}[t]
    \centering
    \includegraphics[width=0.95\linewidth]{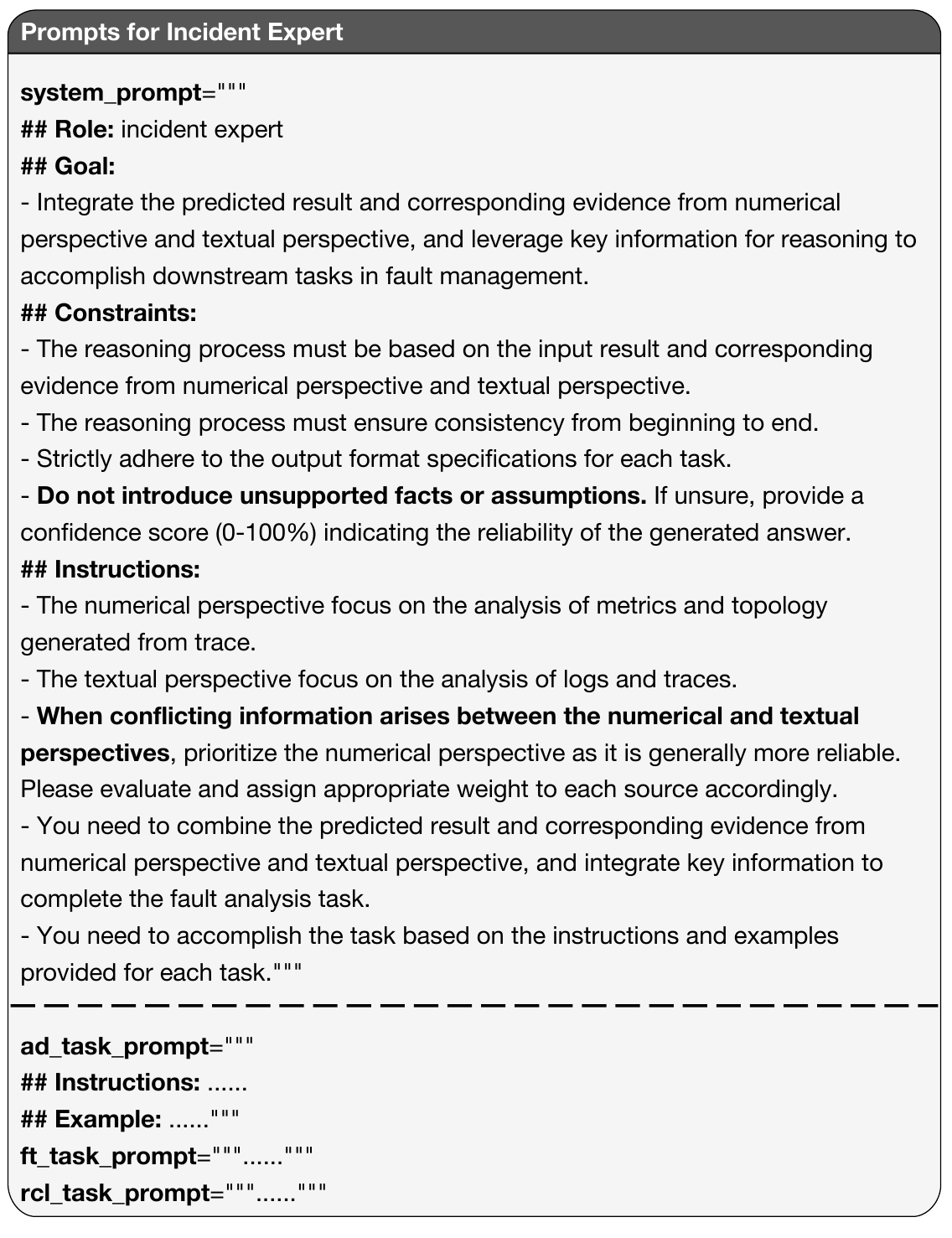}
    \caption{
        The prompt of Incident Expert. This structure prompt is grounded in the ``RGCIE'' principle to mitigate hallucination, providing clear definition of the Incident Expert itself and its corresponding tasks (\ie AD, FT, and RCL). The prompt also contains conflict resolution and aggregation policy to deal with the potential inconsistencies.
    }
    \label{fig:prompt}
    \vspace{-5mm}
\end{figure}

\subsubsection{Coordination Pipeline}

\name adopts the multi-expert collaboration paradigm, leveraging a dispatch-and-aggregation mechanism to fully exploit the specialized strengths of each expert in their respective domains. The pseudo-code for the coordination pipeline is shown in Algorithm~\ref{alg:coordination}.
Specifically, this mechanism first assigns numerical features $\mathcal{F}_{num}$ and textual features $\mathcal{F}_{txt}$ to the Numerical Expert and Textual Expert, respectively, for independent analysis. 
Both experts independently generate results and corresponding reasoning evidence for AD, FT, and RCL.
Subsequently, the Incident Expert performs a comprehensive analysis of the outputs from the Numerical Expert and Textual Expert, balancing information from different sources to resolve potential conflicts and ultimately producing a unified and reliable diagnostic result $\mathcal{R}_{final}$ along with comprehensive reasoning evidence $\mathcal{E}_{final}$.
This dispatch-and-aggregation mechanism not only enables efficient integration of multimodal data but also significantly enhances the accuracy and interpretability of TrioXpert in performing AD, FT, and RCL tasks within complex microservice systems.

\begin{algorithm}[t]
\caption{ Coordination Pipeline in \name}
\label{alg:coordination}
\KwData{
  $\mathcal{F}_{num}$: Numerical feature (from metrics and topology)\\
  $\mathcal{F}_{txt}$: Textual feature (from logs and traces)
}
\KwResult{
  $\mathcal{R}_{final}$: Final diagnostic decision (AD, FT, RCL)\\
  $\mathcal{E}_{final}$: Final explanation reasoning chain
}
\nl $\mathcal{R}_{num}, \mathcal{E}_{num} \gets$ \textbf{NumericalExpert}($\mathcal{F}_{num}$) \tcp*{Process metric-based features}
\nl $\mathcal{R}_{txt}, \mathcal{E}_{txt} \gets$ \textbf{TextualExpert}($\mathcal{F}_{txt}$)  \tcp*{Process logs and traces}
\nl $\mathcal{R}_{final}, \mathcal{E}_{final} \gets$ \textbf{IncidentExpert}($\mathcal{R}_{num}, \mathcal{E}_{num}, \mathcal{R}_{txt}, \mathcal{E}_{txt}$) \tcp*{Aggregate and reconcile}

\Return $(\mathcal{R}_{final}, \mathcal{E}_{final})$
\end{algorithm}

\subsubsection{Conflict Resolution and Aggregation Policy}

Due to potential inconsistencies in incident information across modalities, conflicts may arise between experts. For example, the Numerical Expert may infer a failure type such as ``Node CPU overload'', based on time-series metrics, while the Textual Expert may conclude it is a ``Container Hardware'' issue based on logs and traces. 
To resolve discrepancies through semantic reasoning rather than numerical fusion, we employ a text-guided weighting mechanism. 
Specifically, LLM experts are instructed via natural language prompts to prioritize information sources based on their respective reliability. 

Within the Textual Expert, higher weight is assigned to log entries over trace data. Logs typically contain richer semantic signals, including precise error messages and exception patterns, which are often more indicative of fault semantics than raw trace delays. Furthermore, our keyword-based filtering process ensures that the selected logs $\mathcal{L}$ are contextually aligned with incident symptoms, thus improving their diagnostic quality. 

In Incident Expert, the output of the Numerical Expert is given greater weight, as it extracts features and performs downstream processing based on unfiltered, high-resolution metric data, which tend to be more stable and comprehensive compared to partially filtered logs and trace spans. For example, sustained CPU or memory anomalies are often reliably captured in the metric view, whereas logs may omit such patterns. 
This design has been validated through ablation studies~\ref{tab:Ablation}, which demonstrate that using only the Numerical Expert yields better performance than relying solely on the Textual Expert.

\subsubsection{Hallucination Mitigation Design}
To address the challenge of hallucination in LLMs, \name introduces a comprehensive mitigation strategy centered around refined prompt engineering. Specifically, \name adopts a structured prompt engineering approach grounded in the ``RGCIE'' principle (\ie Role, Goal, Constraints, Instructions, Example) to tailor prompts for different LLMs. By explicitly defining roles, goals, instructions, and constraints, along with providing high-quality examples, this approach promotes strict adherence of LLMs to task requirements, thereby significantly reducing the occurrence of hallucinations. As shown in Fig.~\ref{fig:prompt}, the prompts are further formalized using YAML format, enabling consistent structuring, easy customization, and programmable parsing across diverse model interfaces. This systematic design not only enhances the precision of model outputs but also aligns them more closely with the intended objectives.

Through this strategy, \name effectively mitigates the impact of hallucination, contributing to more robust and reliable performance in incident management tasks.

\section{Experiments}
In this section, we address the following research questions:
\begin{enumerate}[leftmargin=1.5em]
\item \textbf{RQ1:} How does \name perform in AD, FT and RCL?
\item \textbf{RQ2:} Does each component contribute to \name?
\item \textbf{RQ3:} Is \name interpretable and practically actionable for real-world incident management?
\end{enumerate}

\begin{table}[t]
\caption{Detailed information of datasets.}
\label{tab:dataset}
\centering
\resizebox{0.48\textwidth}{!}{%
\begin{tabular}{c|ccccc}
\hline
Datasets & \multicolumn{1}{c}{Instances} & \multicolumn{1}{c}{Failures} & \multicolumn{1}{c}{Normal} & \multicolumn{1}{c}{Failure Types} & \multicolumn{1}{c}{Records} \\ \hline
\multirow{3}{*}{$\mathscr{D}1$}  &    &     &      & & \multicolumn{1}{l}{trace 44,858,388} \\
                                 & 46 & 210 & 3714 & 5   & \multicolumn{1}{l}{log 66,648,685} \\
                                 &    &     &     &   & \multicolumn{1}{l}{metric 20,917,746} \\ \hline
\multirow{3}{*}{$\mathscr{D}2$}  &    &     &      & & \multicolumn{1}{l}{trace 214,337,882} \\
                                 & 18 & 133 & 12297 & 6  & \multicolumn{1}{l}{log 21,356,870} \\
                                 &    &     &      &   & \multicolumn{1}{l}{metric 12,871,809} \\ \hline
\end{tabular}
}

\vspace{-4mm}
\end{table}

\subsection{Experimental Setup}
\subsubsection{Datasets}\label{datasets}

\begin{table*}[htbp]
\centering
\captionsetup{font={large,bf,stretch=1.25},justification=raggedright}
\caption{Performance Comparison on AD, FT, RCL, and Time. ``-'' means this method does not cover the task.}
\renewcommand{\arraystretch}{1.5} 
\resizebox{\textwidth}{!}{%
\begin{tabular}{c|cccccccccc|cccccccccc}
\hline
\multirow{4}{*}[5pt]{Methods} 
                         & \multicolumn{10}{c|}{$\mathscr{D}1$}                                               & \multicolumn{10}{c}{$\mathscr{D}2$}                                               \\ 
                         \cline{2-21} 
                         & \multicolumn{3}{c}{AD}  & \multicolumn{3}{c}{FT}  & \multicolumn{3}{c}{RCL} & \multicolumn{1}{c|}{Efficiency}       & \multicolumn{3}{c}{AD}  & \multicolumn{3}{c}{FT}  & \multicolumn{3}{c}{RCL} & \multicolumn{1}{c}{Efficiency} \\
                         & Precision & Recall & F1 & Precision & Recall & F1 & Top@1 & Top@3  & Avg@5 & Time (s) & Precision & Recall & F1 & Precision & Recall & F1 & Top@1 & Top@3 & Avg@5 & Time (s)\\ 
\cline{1-21}
\name                & 0.880     & 0.972  & \textbf{0.924} & 0.852  & 0.768  & \textbf{0.807} & 0.651  & 0.778 & \textbf{0.773}  & 14.314   &  0.854   & 0.972 & \textbf{0.909} &  0.814 &0.725  & \textbf{0.767}  &  0.550   & 0.775    &  \textbf{0.750}    & 12.597    \\
ART\cite{art}            & 0.759     & 0.621  & 0.683 &  0.786 & 0.794  & 0.790 & 0.683 & 0.762  & 0.757  & 0.872   & 0.593    & 0.972   & 0.737  &  0.860   & 0.650        & 0.740   & 0.375  & 0.825  & 0.738       & 1.363 \\
DiagFusion\cite{diagfusion}  & -      & -     & -     & 0.675  & 0.500  & 0.574 & 0.310  & 0.452  & 0.467      & 4.145   &  -     &   -   & -   &  0.797   & 0.527    &  0.634   &  0.582   & 0.709    & 0.695       & 3.297 \\
Eadro\cite{eadro}        &  0.425     &  0.946 & 0.586    & -   & -   & -  & 0.137   &  0.315    &  0.302     & 0.627   &  0.767   & 0.935     & 0.842   &  -   &  -    &  - &  0.157  & 0.315     & 0.310       & 0.899 \\
\hline
Hades\cite{hades}        &  0.866     &  0.863 & 0.865    & -   & -   & -  & -   &  -    &  -     & 0.104   & 0.867           &0.868        &0.868    &   -        & -       & -   &   -    &        -     & -      & 0.415 \\
MicroCBR\cite{microcbr}        &  -     &  - & -    & 0.667   & 0.796   & 0.726  & -   &  -    &  -     & 0.278   & -           &  -      & -   &  0.629         &  0.678      &  0.653  &  -     &    -         & -      & 0.306 \\
PDiagnose\cite{pdiagnosis}        &  -     &  - & -    & -   & -   & -  & 0.615   &  0.692    &  0.685     & 4.342   &   -        &   -     & -   &  -         &   -     &  -  &  0.037     & 0.296            & 0.285      & 9.919 \\
\hline
\end{tabular}
}
\label{tab:comparison}
\vspace{-4mm}
\end{table*}

To comprehensively evaluate the performance of \name in AD, FT, and RCL, we conducted extensive experiments on two datasets, $\mathscr{D}1$ and $\mathscr{D}2$ (shown in Table~\ref{tab:dataset}). 
These datasets include three modalities of data (\ie metrics, logs, traces), and each incident case was annotated by senior industry experts and researchers with extensive experience in the AIOps community.
Since these datasets not only encompass multimodal information but also provide high-quality annotations consistent with multi-task labels, they support a comprehensive evaluation of \name's performance.
\begin{enumerate}[leftmargin=1.5em]
\item \textbf{$\mathscr{D}1$:} This dataset is derived from a simulated e-commerce system based on a microservice architecture, which is deployed on a top bank's cloud platform.
Its traffic patterns are consistent with real-world business scenarios, and the failure types are summarized from common issues observed in real systems. 
The dataset includes 40 microservice instances and 6 virtual machines, covering five distinct failure types.
\item \textbf{$\mathscr{D}2$:} This dataset is sourced from the management system of a top-tier commercial bank and encompasses real-world business application scenarios. 
The dataset comprises a total of 18 instances, involving various components such as microservices, servers, databases, and Docker, and is categorized into six failure types.
\end{enumerate}

\vspace{-0.3mm}

\subsubsection{Baseline Methods}
We selected six state-of-the-art ~methods as baseline methods. 
Among them, ART \cite{art}, DiagFusion \cite{diagfusion}, and Eadro \cite{eadro} are based on multimodal data analysis and are capable of simultaneously addressing multiple downstream tasks in incident management.
In contrast, Hades \cite{hades}, MicroCBR \cite{microcbr}, and PDiagnose \cite{pdiagnosis} are specifically designed to focus on single tasks, targeting AD, FT, and RCL, respectively.
Due to the absence of the AD module in some methods, we assume that the timestamps of incidents are known during the evaluation of FT and RCL tasks. 
Additionally, we configured the parameters of these baseline methods to align with their original papers. 
For dataset-specific configurations (\eg window length), we made appropriate adjustments based on the characteristics of our data.

\subsubsection{Evaluation Metrics}
As described in Section~\ref{problem}, the goal of \name is to detect failures, determine failure types, and locate the root cause. 
To better reflect the performance of the evaluated methods in real-world applications, we designed and selected different evaluation metrics for each task.

AD and FT are both inherently classification tasks, but their objectives differ: AD is a binary classification task used to determine whether a system has experienced a failure, while FT is a multi-class classification task aimed at identifying the type to which the failure belongs.
To evaluate the performance of these two tasks, we utilized the following metrics: True Positives (TP), False Positives (FP), and False Negatives (FN). 
Based on these metrics, we calculated $Precision=\frac{TP}{TP+FP}$ and $Recall=\frac{TP}{TP+FN}$, and subsequently derived the $F1=\frac{2 \cdot Precision \cdot Recall}{(Precision + Recall)}$ as a comprehensive evaluation criterion.


For the RCL task, we introduced the $Top@K=\frac{1}{N}\sum_{i=1}^N(gt_i \in P_i[1:K])$ metric to calculate the probability that the true root cause is included within the top $K$ predicted results, where $gt$ represents groundtruth, $P$ stands for predictions and $K$ is the total amount of failures.
Additionally, to comprehensively assess model performance, we also employed the $Avg@5=\frac{1}{5}\sum_{i=1}^5Top@K$ metric.


We also include per-case processing time $Time$ (in seconds) to evaluate the runtime efficiency of \name.


\subsubsection{Implementations}
We implemented \name using Python 3.10.16 with PyTorch 2.4.0, Transformers 4.46.0, and DGL 2.4.0+cu124. 
For the backbone LLM, we selected \textit{Qwen2.5-7B-Instruct} due to its state-of-the-art performance on the Open LLM Leaderboard~\cite{openllm} among models under 10B parameters. 
Experiments were conducted on a server with 16-core Intel Xeon Gold 5416S CPU, 376GB RAM, and 8 NVIDIA RTX A6000 GPUs (48GB memory each).
To ensure result reliability, we repeated each experiment five times and reported the average performance.

\subsection{RQ1: Overall Performance}


Table~\ref{tab:comparison} reports the performance of \name compared with six baselines on the $\mathscr{D}1$ and $\mathscr{D}2$ datasets. These baselines include both multi-task frameworks (e.g., ART, DiagFusion, Eadro) and task-specific methods (e.g., Hades for AD, MicroCBR for FT, PDiagnose for RCL).

\name consistently outperforms the baseline methods on both the $\mathscr{D}1$ and $\mathscr{D}2$ datasets, achieving the best performance in AD (improving by 4.7\% to 57.7\%), FT (improving by 2.1\% to 40.6\%), and RCL (improving by 1.6\% to 163.1\%) tasks. 
Specifically, on the $\mathscr{D}1$ and $\mathscr{D}2$ datasets, \name achieves an F1 score exceeding 0.9 in the AD task, an F1 score exceeding 0.75 in the FT task, and an Avg@5 score exceeding 0.75 in the RCL task. 
These results demonstrate that \name, by designing three independent data processing pipelines tailored to the inherent characteristics of different modalities, showcasing significant advantages. 
Furthermore, the experimental results confirm that the textual information in logs and traces indeed contains rich semantic content, which traditional methods may fail to fully exploit.

We also evaluate the runtime efficiency of \name. While the use of LLMs collaborative reasoning inevitably introduces some latency, the average processing time per case remains under 15 seconds across all test scenarios. This level of responsiveness aligns with the latency requirements of most production environments, confirming that \name is suitable for real-time incident management in practical deployments.

\subsection{RQ2: Ablation Study}

\begin{table}[t] 
\centering
\captionsetup{font={large,bf,stretch=1.25},justification=raggedright}
\caption{The evaluation results of ablation study.}
\renewcommand{\arraystretch}{1.5} 
\resizebox{0.45\textwidth}{!}{%
\begin{tabular}{c|ccc|ccc} 
\hline
\multirow{2}{*}{Methods} 
                         & \multicolumn{3}{c|}{$\mathscr{D}1$} & \multicolumn{3}{c}{$\mathscr{D}2$} \\ 
\cline{2-7} 
                         & AD: F1 & FT: F1 & RCL: Avg@5 & AD: F1 & FT: F1 & RCL: Avg@5 \\ 
\cline{1-7}
\name                & \textbf{0.924}  & \textbf{0.807}  & \textbf{0.773}      & \textbf{0.909}  & \textbf{0.767}   & \textbf{0.750}    \\
$\mathcal{A}1$               & 0.725  & 0.190  & 0.667      & 0.832  & 0.685   & 0.625     \\
$\mathcal{A}2$             & nan     & 0.261  & 0.238      & nan     & 0.352   & 0.275      \\
$\mathcal{A}3$                  & 0.672  & 0.398  & 0.534      & 0.583  & 0.284   & 0.608       \\
$\mathcal{A}4$                  & 0.428  & 0.294  & 0.397      & 0.552  & 0.359   & 0.517       \\
$\mathcal{A}5$                  & 0.339  & 0.157  & 0.362      & 0.405  & 0.287   & 0.233       \\
\hline
\end{tabular}
}
\label{tab:Ablation}
\vspace{-4mm}
\end{table}

We conduct ablation studies to assess the contribution of key components in \name. We omit ablation on log/trace filtering due to practical constraints—removing it causes input sequences to exceed LLM context limits, resulting in unacceptable latency. Table~\ref{tab:Ablation} reports the performance of five ablated variants across three tasks on both datasets.

\begin{itemize}[leftmargin=1.5em]
\item $\mathcal{A}1$: Removes the textual pipelines (logs and traces), retaining only metrics and topology features.
\item $\mathcal{A}2$: Removes the numerical pipeline, relying solely on logs and traces.
\item $\mathcal{A}3$: Replaces the multi-expert reasoning with a single LLM.
\item $\mathcal{A}4$: Disables conflict resolution and aggregation; expert outputs are used without coordination.
\item $\mathcal{A}5$: Disables hallucination mitigation by removing structured prompt constraints.
\end{itemize}

\name consistently outperforms all ablated variants on both datasets, demonstrating the necessity of each component. Notably, $\mathcal{A}1$ and $\mathcal{A}2$ suffer from incomplete modality coverage, confirming the value of multimodal input. The AD result of $\mathcal{A}2$ is undefined due to the lack of timestamp granularity in textual data alone, highlighting the importance of numerical signals for temporal localization.

$\mathcal{A}3$ shows a marked drop across all tasks, indicating that the multi-expert collaborative reasoning is more robust than a monolithic LLM under multimodal input load. Similarly, $\mathcal{A}4$ and $\mathcal{A}5$ degrade due to the absence of support mechanisms for resolving expert disagreement and mitigating hallucinations.





\subsection{RQ3: Case Study}
We adopt a qualitative approach to evaluate interpretability, as most baseline methods do not produce explicit reasoning evidence.
Applying quantitative metrics in such a setting would be misleading due to the lack of comparable outputs from baselines.
Importantly, \name has been deployed in Lenovo’s production-grade microservice platform, where it serves as a real-time diagnostic assistant for incident management. 
In traditional operations, identifying the root cause of a complex incident typically requires three experienced OCEs working collaboratively for approximately 2.5 hours, often cycling through five or more hypotheses before locating the true root cause.
By contrast, \name significantly improves both efficiency and accuracy: it requires 26 seconds on average to produce a diagnosis, and identifies the correct root cause within two prediction attempts.
This improvement in efficiency and precision directly translates to reduced TTM and operational cost savings in production environments. 
 We illustrate \name’s interpretability and practical value through three real-world incident cases from Lenovo’s system.

\subsubsection{Case 1: Disk Space Exhaustion}
When users reported 500 errors in a Java application, \name's Numerical Expert detected 100\% disk usage across all service instances, while the Textual Expert identified critical logs: ``ERROR FileSystemError: Failed to write file... No space left on device.''
The Incident Expert synthesized these perspectives, correctly attributing the issue to disk space exhaustion despite the application's multi-replica nature, and explicitly linked the temporary file generation failure to the observed symptoms.

\subsubsection{Case 2: Goroutine Leak}
For an HTTP proxy service with severe latency issues, \name's Numerical Expert revealed steadily increasing CPU utilization correlating with response time degradation. 
The Textual Expert identified the synchronous invocation pattern and absence of coroutine termination.
The Incident Expert connected these observations to diagnose a goroutine leak, explaining how unbounded coroutine creation led to resource exhaustion.

\subsubsection{Case 3: Proxy Misconfiguration}
When a data center worker exhibited widespread timeouts, \name's Numerical Expert detected anomalous network patterns specific to the affected data center. 
The Textual Expert identified the HTTPS\_PROXY environment variable pointing to a foreign data center. 
The Incident Expert reconstructed the faulty request flow, identifying how local requests were routed through a foreign proxy, causing packet loss. 

Two experienced Lenovo OCEs carefully reviewed the reasoning chains across all three cases.
They confirmed that \name's structured, evidence-based reasoning made root causes transparent and traceable, enabling them to quickly verify diagnostic conclusions without repeating manual investigations.
This level of interpretability not only increased their trust in the system’s outputs, but also reduced the effort required for root cause analysis in terms of both time and expert involvement, thereby improving diagnostic efficiency and reducing operational burden in production environments.

\section{Related~work}

\textbf{Methods based on single-modal data}. The metrics, logs, and traces form the core support for the operation and maintenance of modern microservice systems.
In previous research, many scholars have attempted to achieve incident management by conducting in-depth analysis of a specific modality of data.
For example, regarding metrics data, some studies construct dependency graphs to depict the relationships between system components during incidents, with representative methods including CloudRanger \cite{cloudranger}, MS-Rank \cite{msrank}, AutoMAP \cite{automap}, and MicroCause \cite{microcause}. 
For log data, certain approaches \cite{sbld,logcluster,log3c} first perform log parsing and then apply hierarchical agglomerative clustering to cluster incidents.
As for trace data, many methods \cite{tracerank,trank,microrank} typically use the 3-sigma approach to determine whether delays are abnormal for anomaly detection, followed by spectrum-based methods for root cause localization.
However, these approaches rely exclusively on one modality and fail to capture cross-modal cues. When anomalies manifest differently across modalities—or are absent from the chosen one—such methods often break down.

\textbf{Methods based on multimodal data}. In recent years, researchers have increasingly recognized that analyzing data from multiple modalities can facilitate a more comprehensive sketch of the operational status of microservice systems. 
Consequently, many studies have attempted to integrate metrics, logs, and trace data, including ART \cite{art}, DiagFusion \cite{diagfusion}, Eadro \cite{eadro}, and DeepHunt \cite{deephunt}. However, since log and trace data are typically presented in the form of massive text, traditional processing approaches tend to extract statistical features while primarily utilizing the topological structure of traces. 
For instance, ART constructs a unified time-series matrix by calculating the frequency of log templates over a specific time period and the distribution of different status codes within traces. 
Although these methods can achieve multimodal data fusion to some extent, they inevitably result in significant information loss, particularly failing to fully leverage the textual information contained in logs and traces. 

\textbf{Methods based on Large Language Models}. With the emergence of powerful natural language understanding and reasoning capabilities in LLMs, researchers have explored their application in the field of incident management. 
However, existing works have yet to adequately address the key limitations in the current incident management domain. 
For instance, Oasis\cite{oasis} is primarily used for automatic incident summary generation, Xpert\cite{xpert} focuses on generating KQL query languages, and Nissist\cite{nissist} is dedicated to recommending failure remediation measures. 
None of these methods directly cover multiple stages of the incident management lifecycle. 
Furthermore, a series of works \cite{icl_rca,rcacopilot,comet} have analyzed the root cause and failure type, but they rely solely on diagnostic information of incidents, failing to fully leverage multimodal data such as metrics, logs, and traces.
More critically, these systems do not address core limitations of LLMs—especially hallucination and context window constraints—which hinder their reliability in high-stakes operational environments.

\section{Discussion}
\subsection{Limitations and Possible Solutions}

Despite the effectiveness of \name, several limitations remain.
(1) Generalization of numerical modeling. The first pipeline employs self-supervised learning over metric features, which limits generalizability across evolving service topologies or unseen deployment environments. Frequent retraining is required to adapt, impeding deployment scalability. Incorporating training-free anomaly detectors, such as PatternMatcher~\cite{patternmatcher}, may reduce this retraining overhead while preserving diagnostic sensitivity.
(2) Expert coordination and robustness. \name adopts a dispatch-and-aggregation multi-expert collaboration architecture, which assumes that each expert (numerical, textual) produces independently correct outputs. In practice, error propagation can occur due to inconsistent reasoning or incomplete evidence. Future work may introduce an adaptive correction mechanism (e.g., a Judge LLM) that verifies cross-expert consistency and triggers targeted re-analysis when semantic contradictions are detected.

\subsection{Threats to Validity}

We recognize the following threats to the validity of our results:
(1) Dataset scope and system complexity. Both $\mathscr{D}1$ and $\mathscr{D}2$ are derived from controlled environments with fewer microservices and lower workload volatility than industrial systems. Although representative, they may not capture failure dynamics at scale. Extending evaluations to real-world datasets will be essential.
(2) Simplified failure cases. The failure patterns in $\mathscr{D}1$ and $\mathscr{D}2$ are relatively simple and do not fully capture the complexity of real-world industrial incidents, such as cascading faults or timing-sensitive interactions. This may limit \name's generalizability to more intricate failure scenarios. Nonetheless, the current setup demonstrates strong results with a 7B-scale LLM. Scaling to larger models and more expressive inputs offers a promising direction for improving robustness in production environments.

\subsection{Efficiency and Impact on Model Size}

\name achieves low-latency inference despite integrating LLM-based summarization and collaborative reasoning. Across all test cases, the average end-to-end diagnosis is completed within 15 seconds, ensuring no adverse impact on TTM and satisfying industrial requirements for rapid incident response.
Regarding model size, \name currently uses 7B-parameter LLMs for each expert module. This configuration, combined with carefully engineered prompts, achieves strong performance on both $\mathscr{D}1$ and $\mathscr{D}2$, while effectively mitigating hallucinations and preserving interpretability. Although larger models may offer marginal performance gains, the increased computational overhead and latency are not justified under current task complexity.
That said, in future deployments targeting more intricate failure scenarios, scaling up the model size may become necessary. At present, the 7B setup offers a favorable trade-off between performance, interpretability, and operational efficiency.

\section{Conclusion}

This work introduces \name, an end-to-end framework for unified incident management in microservice systems through modality-specific preprocessing and collaborative LLM reasoning. 
By preserving semantic richness while addressing LLM limitations via structured prompts, it simultaneously achieves significant accuracy improvements across AD, FT, and RCL.
The approach shows that modality-aware representations and structured LLM collaboration enable scalable, interpretable incident management.
While designed for microservices, its principles can generalize to other complex, high-stakes systems requiring multisource analysis.

\section{Acknowledgement}
This work is supported by the Advanced Research Project of China (31511010501), the National Natural Science Foundation of China (62272249, 62302244), the Fundamental Research Funds for the Central Universities (XXX-63253249), and the CCF-Lenovo Blue Ocean Research Fund.

\bibliographystyle{IEEEtran}
\bibliography{main.bib}

\end{document}